\documentclass[conference]{IEEEtran}

\usepackage{ifpdf}
\ifpdf
\usepackage[final,pdfborder={0 0 0}]{hyperref}
\hypersetup{pdfauthor={Georges Kaddoum, Pascal Giard},%
  pdftitle={Analog Network Coding for Multi-User Spread-Spectrum Communication Systems},
  colorlinks=true,
  linkcolor=blue,
  citecolor=blue
}%
\fi

\usepackage{cite}
\usepackage[cmex10]{amsmath}
\usepackage{graphicx}
\usepackage{multicol}
\usepackage{amssymb}
\usepackage{epsfig}
\usepackage[cmex10]{amsmath}

\usepackage{flushend} %

\usepackage{array}
\newcolumntype{C}[1]{>{\centering\arraybackslash}m{#1}}

\usepackage{color}

\newcommand\blfootnote[1]{%
  \begingroup
  \renewcommand\thefootnote{}\footnote{#1}%
  \addtocounter{footnote}{-1}%
  \endgroup}

\usepackage{tikz,pgfplots}
\pgfplotsset{compat=newest} 
\pgfplotsset{plot coordinates/math parser=false}
\usetikzlibrary{plotmarks,shapes.geometric,arrows}
\tikzset{
    >=stealth'
}

\begin{document}

\title{Analog Network Coding for Multi-User Spread-Spectrum Communication Systems}

\author{\authorblockN{Georges Kaddoum\authorrefmark{1}, \textit{Member, IEEE} and Pascal Giard\authorrefmark{2}\authorrefmark{1}, \textit{Student Member, IEEE}}\\

\authorblockA{\authorrefmark{1}LaCIME, Department of Electrical Engineering, \'{E}cole de technologie sup\'{e}rieure.\\
\authorblockA{\authorrefmark{2}ISIP, Department of Electrical and Computer Engineering, McGill University.\\
Emails: georges.kaddoum@lacime.etsmtl.ca, pascal.giard@mail.mcgill.ca}}\\
}

\maketitle

\blfootnote{This work is partially supported by the discovery grant $435243-2013$ from the Natural Research Council of Canada (NSERC).}

\begin{abstract}
This work presents another look at an analog network coding scheme for multi-user spread-spectrum communication systems. Our proposed system combines coding  and cooperation between a relay and users to boost the throughput and to exploit interference. To this end, each pair of users, $\mathcal{A}$ and $\mathcal{B}$, that communicate with each other via a relay $\mathcal{R}$ shares the same spreading code. The relay has two roles, it synchronizes network transmissions and it broadcasts the combined signals received from users. From user $\mathcal{B}$'s point of view, the signal is decoded, and then, the data transmitted by user $\mathcal{A}$ is recovered by subtracting user $\mathcal{B}$'s own data. We derive the analytical performance of this system for an additive white Gaussian noise channel with the presence of multi-user interference, and we confirm its accuracy by simulation.
\end{abstract}

\IEEEpeerreviewmaketitle

\section{Introduction}

Dealing with interference when signals from multiple nodes arrive simultaneously is a major challenge in wireless networks. Recently, many researchers showed a high interest in using network coding techniques like Analog Network Coding (ANC) or more generally Physical Layer Network Coding (PNC) to mitigate the effect of interference and to boost the network throughput~\cite{soun11,Katti07,Zhan06}.

In a PNC communication scheme, a pair of users $\mathcal{A}$ and $\mathcal{B}$ (\textit{i.e.} terminal nodes) first simultaneously transmit their data signals to a relay. The relay then transforms the received overlapped signals into a network-coded message (\textit{e.g.} does a bit-wise XOR of the messages from each terminal node) and broadcasts the resulting signal. By knowing the reversible network-coding operation carried out by the relay, each terminal node is able to extract the message sent by the other node of its pair by removing its own contribution from the received signal. Thus, two time slots are required for both users $\mathcal{A}$ and $\mathcal{B}$ to exchange one message. This constitutes an improvement over the traditional relaying scheme where four time slots are needed \cite{Zhan06} to achieve the same goal.

On the other hand, in ANC schemes as proposed in \cite{Katti07}, a relay simply forwards superimposed signals to users. Signal decoding is carried out at the user level since the linear mapping is naturally done in the physical channel. An ANC scheme can be realized and implemented with ease. However, as the relay does not remove the noise, it is forwarded along with the desired signals. Despite this disadvantage, an ANC scheme -- with some extensions to allow multi-user transmission and to exploit the interference present in the network -- is preferred since a classic PNC scheme remains significantly more complex. In a PNC scheme, the following supplementary operations would be executed on a relay node: decode the received signals that use different spreading codes, map the resulting signals, apply the appropriate spreading code depending on the destination, and finally forward the signals.

The proposed ANC scheme in this paper can be used for a spread-spectrum communication system using any spread-spectrum code (Gold, Kasami, Walsh, chaotic, ...).  The choice of the chaotic signals is attributable to the advantages offered by this class of spreading sequences~\cite{Rah13,Vitali2006}. Notably, their sensitivity to initial conditions allows chaotic maps to theoretically generate an infinite number of very low cross-correlated signals. These wideband signals have been shown to be well-suited for multi-user spread-spectrum (SS) modulation applications~\cite{Lau03,Kur05,Val12}. In addition, some types of chaotic modulation carry the same advantages as other conventional SS modulations, including mitigation of fading channels~\cite{Xia04,Ren13}, jamming resistance, low probability of interception (LPI)~\cite{Yu2005} and secure communications~\cite{6179320}. Furthermore, many papers demonstrated that chaos-based sequences produce good results compared to that of Gold sequences or other independent and identically distributed sequences, notably reducing multi-user interference and peak-to-average power ratio (PAPR)~\cite{Rah13,Vitali2006}. Moreover, code synchronization techniques used in traditional SS communication systems to achieve robust synchronization in two distinct phases (\textit{i.e.} code acquisition and tracking) were recently applied to chaos SS communication systems~\cite{Kad08s, Val12}. In this paper, we assume that this code synchronization technique is adequate to demodulate the transmitted symbols in our system.

In the literature, many chaos-based SS systems with coherent and non-coherent receivers are proposed and evaluated \cite{6198019}. The most noteworthy systems include Chaos Shift Keying (CSK)~\cite{Lau03}, Differential Chaos Shift Keying (DCSK)~\cite{Kol96, 5871787, 6184294} and mutli-carrier DCSK~\cite{6560492}. With coherent receivers, like in CSK systems, a chaotic signal is used as a spreading sequence at the transmitter to spread the information data signal. At the receiver, chaotic synchronization is required in order to demodulate the transmitted bits~\cite{Kad08s,5743054,Val12}; as well as a coherent correlator and a threshold detector are used to decode the signal. On the other hand, when DCSK is used, chaotic synchronization is not required at the receiver \textit{i.e.} exact knowledge of the chaotic signal is not mandatory in order to demodulate the transmitted bits. However, a DCSK system is less secure, and does not perform as well as a coherent chaos-based communication system \cite{Lau03}.

\subsubsection*{Related Works} In \cite{Kulhandjian12,Shu12}, ANC schemes for CDMA systems were proposed where each user in the system is assigned one distinct spreading sequence. To decode a message from a given user in the network, a node applies the spreading sequence of the other user. In other words, the proposed scheme requires a priori knowledge of the list of all users along with their associated spreading sequences in order to decode the transmitted data without exploiting multi-user interference. In this paper, we propose a different approach for SS systems. In our network, we consider a two-way wireless system with $L$ source nodes (\textit{i.e.} users) ($L$ is an even number for simplicity) with one single relay. Each node uses direct-sequence SS modulation. The main aim of this paper is to propose an ANC scheme for multi-user SS systems which reduces the number of time slots required to exchange data, to exploit the interference, and to increase the throughput of the network.  To tackle this challenge, we propose a new design in which each pair of users wishing to communicate in this network via the relay share a spreading code. The fact that both users of a pair use the same spreading code exploits the interference and boosts the throughput of the system.

In an ANC scheme, network synchronization is critical to any reliable implementation. In our design, network synchronization is maintained by the relay. The relay sends a request-to-transmit signal once it is ready to receive the nodes' signals. This operation forces users to simultaneously transmit their messages. Once the users receive the request-to-transmit signal, they start their transmission phase in the first time slot. In that phase, the $L$  nodes transmit their signals simultaneously to the relay. In the second time slot, or during the broadcasting phase, the relay retransmits the summed received signals to all users in the network. Each user first multiplies the received signal with its local spreading sequence, it then  decodes the signal, and finally, it removes its own data in order to recover the data sent by the other member of the pair. These operations are further detailed later in this paper.

\subsubsection*{Paper Outline} The remainder of this paper is organized as follows. In Section~\ref{sect:system}, we briefly present the chaos-based spread-spectrum communication system. Section~\ref{sect:scheme} is dedicated to explaining the design of the proposed ANC-SS scheme. The analytical performance analysis is presented in Section~\ref{sect:perf} where we derive the analytical bit error rate (BER) expressions for an additive white Gaussian noise (AWGN) channel in presence of multi-user interference. Simulation results validating our analytical approach are shown in Section~\ref{sect:sim}, and finally, some concluding remarks are given.

\section{CSK Communication System}
\label{sect:system}

The studied spread-spectrum system uses a CSK modulation in which data information symbols $(s_{i}= \pm 1)$ with period $T_s$ are spread by a chaotic sequence $x_{k}$. A new chaotic sample (or chip) is generated at time intervals equal to $T_{c}$ ($x_{k}=x(kT_{c})$). The emitted signal $e (t)$ at the output of a transmitter is:
\begin{equation}
e (t) = \sum\limits_{i = 1}^\infty {\sum\limits_{k = 1}^{\beta} {s_i x_{i\beta + k} } } g(t - (i\beta + k)T_c )
\end{equation}
where $\beta$ is the spreading factor that is equal to the number of chaotic samples in a symbol duration $(\beta=T_s/T_c)$; and $g(t)$ is the pulse shaping filter. A rectangular pulse of unit amplitude on $[0, \, T_c]$ is used in this paper.  

For an AWGN channel, the received signal is:
\begin{equation}
u(t) = e(t)+ n(t)
\end{equation}
where $n(t)$ is additive white Gaussian noise with zero mean and a power spectral density equal to $N_{0}/2$. 
In order to demodulate the transmitted bits, the received signal is first multiplied by the local chaotic sequence, and then integrated over a symbol duration $T_s$. Finally, the transmitted bits are estimated by computing the sign of the decision variable at the output of the correlator.
\begin{equation}
Ds_i = {\rm sign}\left( {s_i T_c \sum\limits_{k = 1}^{\beta } {\left( {x_{i\beta + k} } \right)^2 } + w_i = s_i E_b^{(i)} + w_i } \right)
\end{equation}
where ${\rm sign(.)}$ is the sign operator, $E_b^{(i)}$ is the bit energy of the $i^{th}$ bit and $w_i$ is the noise after despreading and integration. For the sake of simplicity, we omit the pulse shaping filter expression and we normalize the time chip $T_c$ to $1$.

Throughout the paper, a Bernoulli mapping function is used 
\begin{equation}\label{cpf}
x_{k+1}=
      \left\{\begin{array}{l}
	Gx_k-F~~\textup{if}~x_k\geq0\\
	Gx_k+F~~\textup{otherwise}
      \end{array}\right.
    \end{equation}
where $F$ and $G$ are constants. This map is chosen to generate the chaotic sequences because of its implementation efficiency~\cite{Giard2012}. In addition, chaotic sequences are normalized so that ${\rm E}[x_{k}^2]=1 $ and have a zero mean to avoid DC transmission \textit{i.e.} ${\rm E}[x_{k}]=0$. 

Generally, two different approaches have been adopted to use chaotic signals in digital communication systems. Either the real value of a chaotic signal is used to modulate the bits to be transmitted~\cite{Lau03} or the chaotic signal is quantized and used to transmit data \cite{Kur05,4806056,6486530}. The latter offers better performance than conventional spreading systems. However, that approach leads to a loss of chaotic signal properties \cite{Rah13}. Thus, it is the first approach that is adopted in our paper. Note that our general analysis can nonetheless be adapted to quantized chaotic signals.

\section{ANC-SS Communication Scheme}
\label{sect:scheme}

In this section, $s_{\mathcal{A}}$, $s_{\mathcal{B}}$, and $s_{\mathcal{R}}$ refer to the data of users $\mathcal{A}$, $\mathcal{B}$, and the mapped data in the relay $\mathcal{R}$ respectively represented by the signals $e_{\mathcal{A}}$, $e_{\mathcal{B}}$, and $e_{\mathcal{R}}$. The use of an ANC scheme limits the required number of time slots to two. That represents an improvement in comparison to existing non-network coded schemes or straightforward network coding schemes~\cite{Zhan06}.

In an ANC scheme, $s_{\mathcal{R}}$ is computed directly from $s_{\mathcal{A}} + s_{\mathcal{B}}$ which occurs naturally in the physical channel from the superimposition of signals. In this case, the arithmetic sum is considered a form of network coding. In conventional multi-user spread-spectrum systems each user has its own spreading code. In the context of ANC schemes, a key requirement is that each node must have the knowledge of all spreading sequences attributed to users. This coding does not allow exploition of interference. The main challenge in our paper is to provide a reliable scheme that is easy to implement while mitigating the interference present in the network.

\begin{table*}[htb!]
  \small
  \caption{ANC-SS mapping scheme}
  \centering  
  \begin{tabular}{ | C{2.5cm} | C{2.5cm} | C{3.5cm} | C{3.5cm} | C{3.5cm} |}
    \hline
    Spreading symbol from $\mathcal{A}$    & Spreading symbol from $\mathcal{B}$  &   Signal broadcast from the relay  & Despreading/decoding symbol at $\mathcal{B}$ & Extracting symbols of $\mathcal{B}$, recovering symbols of $\mathcal{A}$ \\ \hline\hline
    $1x_{(\mathcal{A},\mathcal{B})}$    &   $1x_{(\mathcal{A},\mathcal{B})}$   &  $2x_{(\mathcal{A},\mathcal{B})}$&  2   & $(2-1)=1$ \\ \hline
    $1x_{(\mathcal{A},\mathcal{B})}$   &  $-1x_{(\mathcal{A},\mathcal{B})}$   & $0x_{(\mathcal{A},\mathcal{B})}$ &  0   & $(0-(-1))=1$  \\ \hline
    $-1x_{(\mathcal{A},\mathcal{B})}$ &  $1x_{(\mathcal{A},\mathcal{B})}$    & $0x_{(\mathcal{A},\mathcal{B})}$ &  0   & $(0 -1)=-1$  \\ \hline
    $-1x_{(\mathcal{A},\mathcal{B})}$ &  $-1x_{(\mathcal{A},\mathcal{B})}$   & $-2x_{(\mathcal{A},\mathcal{B})}$&  -2 & $(-2-(-1))=-1$  \\ \hline
  \end{tabular}
  \label{map-ss}
\end{table*}

In our proposed scheme, each pair of users $\mathcal{A}$ and $\mathcal{B}$ wishing to communicate with each other must use the same spreading code $x_{(\mathcal{A},\mathcal{B})}$. This scheme is illustrated in Fig.~\ref{netw} and summarized in Table~\ref{map-ss}. This particular design allows the exploitation of interference and users can still extract the information from the other member of the pair. The detailed presentation of our communication scheme along with the relevant mathematical descriptions can be broken down into the following four phases in which two time slots are allocated for the transmission and broadcasting phases respectively:

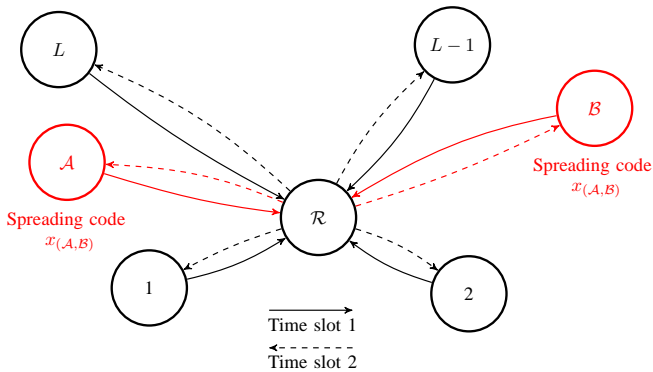
\begin{figure}[htb!]
\centering
\resizebox{9cm}{!}{
  \begin{tikzpicture}

[every node/.style={inner sep=0pt}]
\node (2) [circle, minimum size=40.0pt, fill=white, line width=1.25pt, draw=black] at (213.125pt, -200.625pt) {\textcolor{black}{$\mathcal{R}$}};
\node (3) [circle, minimum size=40.0pt, fill=white, line width=1.25pt, draw=red] at (360.0pt, -142.5pt) {\textcolor{red}{$\mathcal{B}$}};
\node (4) [circle, minimum size=40.0pt, fill=white, line width=1.25pt, draw=black] at (75.0pt, -111.25pt) {\textcolor{black}{$L$}};
\node (5) [circle, minimum size=40.0pt, fill=white, line width=1.25pt, draw=black] at (284.375pt, -108.75pt) {\textcolor{black}{$L-1$}};
\node (1) [circle, minimum size=40.0pt, fill=white, line width=1.25pt, draw=red] at (79.375pt, -171.25pt) {\textcolor{red}{$\mathcal{A}$}};
\node (7) [circle, minimum size=40.0pt, fill=white, line width=1.25pt, draw=black] at (123.125pt, -238.125pt) {\textcolor{black}{1}};
\node (6) [circle, minimum size=40.0pt, fill=white, line width=1.25pt, draw=black] at (293.125pt, -241.25pt) {\textcolor{black}{2}};

%% Use ellipses instead?
%% \node (2) [ellipse, minimum height=20.0pt, minimum width=40.0pt, fill=white, line width=1.25pt, draw=black] at (213.125pt, -200.625pt) {\textcolor{black}{$\mathcal{R}$}};
%% \node (3) [ellipse, minimum height=20.0pt, minimum width=40.0pt, fill=white, line width=1.25pt, draw=red] at (360.0pt, -142.5pt) {\textcolor{red}{$\mathcal{B}$}};
%% \node (4) [ellipse, minimum height=20.0pt, minimum width=40.0pt, fill=white, line width=1.25pt, draw=black] at (75.0pt, -111.25pt) {\textcolor{black}{$L$}};
%% \node (5) [ellipse, minimum height=20.0pt, minimum width=40.0pt, fill=white, line width=1.25pt, draw=black] at (284.375pt, -108.75pt) {\textcolor{black}{$L-1$}};
%% \node (1) [ellipse, minimum height=20.0pt, minimum width=40.0pt, fill=white, line width=1.25pt, draw=red] at (79.375pt, -171.25pt) {\textcolor{red}{$\mathcal{A}$}};
%% \node (7) [ellipse, minimum height=20.0pt, minimum width=40.0pt, fill=white, line width=1.25pt, draw=black] at (123.125pt, -238.125pt) {\textcolor{black}{1}};
%% \node (6) [ellipse, minimum height=20.0pt, minimum width=40.0pt, fill=white, line width=1.25pt, draw=black] at (293.125pt, -241.25pt) {\textcolor{black}{2}};

\node (8) [circle, minimum size=10.0pt, fill=white, line width=1.25pt] at (181.25pt, -250.0pt) {};
\node (9) [circle, minimum size=10.0pt, fill=white, line width=1.25pt] at (236.875pt, -250.0pt) {};
\node (10) [circle, minimum size=10.0pt, fill=white, line width=1.25pt] at (181.25pt, -270.0pt) {};
\node (11) [circle, minimum size=10.0pt, fill=white, line width=1.25pt] at (236.875pt, -270.0pt) {};

\draw [line width=0.625, ->, dashed, color=red] (2) to  [in=206, out=17] (3);
\draw [line width=0.625, ->, dashed, color=red] (2) to  [in=357, out=157] (1);
\draw [line width=0.625, ->, dashed, color=black] (2) to  [in=337, out=137] (4);
\draw [line width=0.625, ->, dashed, color=black] (2) to  [in=143, out=343] (6);
\draw [line width=0.625, ->, dashed, color=black] (2) to  [in=223, out=62] (5);
\draw [line width=0.625, ->, dashed, color=black] (2) to  [in=28, out=197] (7);
\draw [line width=0.625, ->, color=red] (3) to  [in=31, out=191] (2);
\draw [line width=0.625, ->, color=red] (1) to  [in=173, out=343] (2);
\draw [line width=0.625, ->, color=black] (6) to  [in=323, out=163] (2);
\draw [line width=0.625, ->, color=black] (7) to  [in=212, out=13] (2);
\draw [line width=0.625, ->, color=black] (5) to  [in=43, out=242] (2);
\draw [line width=0.625, ->, color=black] (4) to  [in=152, out=322] (2);
\node at (78.75pt, -202.5pt) {\textcolor{red}{Spreading code}};
\node at (81pt, -214.75pt) {\textcolor{red}{$x_{(\mathcal{A},\mathcal{B})}$}};
\node at (358.75pt, -173.75pt) {\textcolor{red}{Spreading code}};
\node at (361pt, -186pt) {\textcolor{red}{$x_{(\mathcal{A},\mathcal{B})}$}};

\draw [line width=0.625, ->, color=black] (8) to (9);
\node at (210.0pt, -257.5pt) {\textcolor{black}{Time slot 1}};
\draw [line width=0.625, ->, dashed, color=black] (11) to (10);
\node at (210.0pt, -277.5pt) {\textcolor{black}{Time slot 2}};

\end{tikzpicture}
}
\caption {ANC-SS scheme with $L$ users.}
\label{netw}
\end{figure}

\subsubsection{Request-to-transmit phase} In this preliminary phase, the relay broadcasts a signal informing all users to transmit their signals to the relay. This operation allows the synchronization of the transmission taking place in the first time slot. 

\subsubsection{Transmission phase} In this first time slot, the nodes transmit their signals to the relay. Each pair of users  which want to communicate between them use the same spreading code. The received signal $u_N (t)$ at the relay $\mathcal{R}$ is:
\begin{equation}\label{up_sig}
\begin{array}{l}
 u_N (t) = \sum\limits_{i = 1}^\infty  {\sum\limits_{k = 1}^\beta  {\left( {s_{\mathcal{A},i}  + s_{\mathcal{B},i} } \right)} } x_{(\mathcal{A},\mathcal{B}),i\beta  + k}  \\ 
  + \sum\limits_{i = 1}^\infty {\sum\limits_{\substack{
        \scriptstyle l \in \mathcal{E} \\
        \scriptstyle l \ne m}} \sum\limits_{\scriptstyle m \in \mathcal{E}} {\sum\limits_{k = 1}^\beta  {\left( {s_{l,i}  + s_{m,i} } \right)x_{(l, m),i\beta  + k} } }  + n_N (t)}  \\ 
 \end{array}
\end{equation}
where $s_{l,i} $ is the $i^{th}$ symbol of user $l$ spread by the sequence $x_{(l, m),i\beta  + k} $ and $\mathcal{E}=\left\{1,2,..,L\right\}\backslash\{\mathcal{A},\mathcal{B}\}$. In \eqref{up_sig}, $ u_N (t)$ is the received signal coming from different nodes. The first term in this equation represents the sum of the transmitted signals from the user pair $\mathcal{A}$ and $\mathcal{B}$. The second term in \eqref{up_sig} represents the multi-user interference signal in which each pair of users $l$ and $m$ use the same spreading code $x_{(l,m),i\beta  + k}$, and  $n_N (t)$ represents the additive white  Gaussian noise of the transmission phase with zero mean and variance equal to $N_0/2$.

\subsubsection{Broadcasting phase} In the second time slot, the relay forwards the received signal $u_N (t)$ to the different nodes in the network. In our system we assume that the relay just forwards the signal without amplifying it. Furthermore, the  amplifying operation can be added to the relay when it is in the presence of a fading channel. This can be done without any modification to our ANC-based design. Thus, the signal broadcast by the relay $u_{\mathcal{R}} (t)$ is:
\begin{equation}
u_{\mathcal{R}} (t) = u_N (t) + n_{\mathcal{R}} (t)
\end{equation}
where  $n_{\mathcal{R}} (t)$ is the additive white Gaussian noise of the broadcasting phase with zero mean and variance equal to $N_0/2$.

\subsubsection{Decoding phase} For user $\mathcal{B}$ to be able to extract the information from the other member of the pair, $\mathcal{A}$, the received signal is first despread using the same spreading code $x_{(\mathcal{A},\mathcal{B}),i\beta  + k}$ and then decoded. Lastly, node $\mathcal{B}$ removes its own data from the decoded signal, as detailed in Table~\ref{map-ss}, to recover the data originating from node $\mathcal{A}$. 

\section{Performance Analysis of the ANC-SS System}
\label{sect:perf}

The second objective of this paper is to validate the concept of the proposed scheme. Once the system is explained, the performance of the ANC-SS system over an AWGN channel and in presence of multi-user interference is investigated. 
As mentioned in Section~\ref{sect:scheme}, in the first phase, users $\mathcal{A}$ and $\mathcal{B}$ transmit their signals which are later forwarded by the relay. In the reception phase, the received signal is the combination of the two transmitted signals from $\mathcal{A}$ and $\mathcal{B}$ along with the additive noise signals $n_N (t)$ and $n_{\mathcal{R}} (t)$, and the multi-user interference.  The decision variable for a given bit $i$ transmitted from $\mathcal{A}$ is:
\begin{equation} \label{dec_var12}
D_{s_{\mathcal{A},i} }  =  \sum\limits_{k = 1}^\beta  {u_{\mathcal{R}} (t)} x_{(\mathcal{A},\mathcal{B}),i\beta  + k}\;.
\end{equation}

The mean of $D_{s_{\mathcal{A},i} }$ is reduced to the term given in \eqref{m1} because the elements of the noise components and chaotic signals are independent and both have a zero mean:
\begin{equation}\label{m1}
{\rm E}\left[ D_{s_{\mathcal{A},i} } \right] = E_b^{(i)} s_{\mathcal{R},i}
\end{equation}
where $s_{\mathcal{R},i}=s_{\mathcal{A},i}+s_{\mathcal{B},i} \in \lbrace -2, 0, 2 \rbrace$, and  $E_b^{(i)}  = \sum\limits_{k = 1}^\beta  {x_{(\mathcal{A},\mathcal{B}),i\beta  + k}^2 } $.

The signal given in \eqref{dec_var12} is decoded after comparing its amplitude to two  predetermined thresholds $\gamma_1$ and $\gamma_2$. Based on the model provided in \cite{Pro95}, the optimal thresholds in a noise free environment using the values given in \eqref{m1} are:
\begin{equation}
\begin{array}{l}
 \gamma_1  = \left( {\frac{{0-2E_b }}{2}} \right) = - E_b \\ 
 \gamma_2  = \left( {\frac{{ 2E_b+ 0}}{2}} \right) =   E_b \\ 
 \end{array}
\end{equation}
where $E_b  = \sum\limits_{k = 1}^\beta  {x_{(\mathcal{A},\mathcal{B}),\beta  + k}^2 }\;$.

In fact, when the received signal amplitude of \eqref{dec_var12} is grater than  $\gamma _2$, we decide that the symbol is $2$. Similarly, if the received signal amplitude is lower than  $\gamma _1$, we decide that the symbol is $-2$. Finally, if the signal amplitude is between   $\gamma _1$ and  $\gamma _2$, it is declared $0$. According to Table~\ref{map-ss}, the decoded signal space is $ \lbrace -2, 0, 2 \rbrace$ with corresponding probabilities of $25 \%$,  $50 \%$ and $25 \%$ respectively.

Finally, the estimated  data of user $\mathcal{A}$  $\hat {s}_{\mathcal{A}}$ is computed  by subtracting the data symbol $s_{\mathcal{B}}$  from the decoded symbol obtained from the  signal given in \eqref{dec_var12}. According to the central limit theorem, multi-user interference follows a Gaussian distribution. Therefore, the BER of $ \hat {s}_{\mathcal{A}}$ based on the signal space probabilities is the BER of the decoded signal given as follows:
\begin{equation} \label{ber_e}
\small
\begin{array}{l}
 BER =  \\ 
 \frac{1}{{2\sqrt{2\pi}  \sigma _{D_{s_{\mathcal{A}} } } }}\left( {\int\limits_{ - \infty }^{\gamma _1 } {\exp \left( {\frac{{ - r^2 }}{{2\sigma _{D_{s_{\mathcal{A}} } }^2 }}} \right)dr}  + \int\limits_{\gamma _2 }^\infty  {\exp \left( {\frac{{ - r^2 }}{{2\sigma _{D_{s_{\mathcal{A}} } }^2 }}} \right)dr} } \right) + \\ 
  \frac{1}{{4\sqrt{2\pi}  \sigma _{D_{s_{\mathcal{A}} } } }} \hspace{-0.1cm} \left( \hspace{-0.1cm} {\int\limits_{\gamma _1 }^{\gamma _2 } {\exp \left( {\frac{{ - \left( {r + 2E_b} \right)^2 }}{{2\sigma _{D_{s_{\mathcal{A}} } }^2 }}} \right)dr}  + \int\limits_{\gamma _1 }^{\gamma _2 } {\exp \left( {\frac{{ - \left( {r - 2E_b} \right)^2 }}{{2\sigma _{D_{s_{\mathcal{A}} } }^2 }}} \right)dr} }\hspace{-0.1cm} \right).
 \end{array}
\end{equation}

\eqref{ber_e} requires the computation of the decision variable variance $\sigma^2_{D_{s_{\mathcal{A}} } }$. The development of \eqref{dec_var12} for a given bit $i$ is:
\begin{equation}\label{dec_var1}
\begin{array}{l}
 D_{S_{{\mathcal{A}},i} }  =  \\ 
 \begin{array}{l}
 \sum\limits_{i = 1}^\infty  {\sum\limits_{k = 1}^\beta  {\left( {s_{{\mathcal{A}},i}  + s_{{\mathcal{B}},i} } \right)} } x_{(\mathcal{A},\mathcal{B}),i\beta  + k}^2  + \sum\limits_{k = 1}^\beta  {n_{N,k} x_{(\mathcal{A},\mathcal{B}),i\beta  + k}^{} }  \\ 
  + \sum\limits_{i = 1}^\infty  {\sum\limits_{
        \substack{\scriptstyle l \in \mathcal{E}\\
        l \ne m}} {\sum\limits_{m \in \mathcal{E}} {\sum\limits_{k = 1}^\beta  {\left( {s_{l,i}  + s_{m,i} } \right)x_{(l,m),i\beta  + k}^{} x_{(\mathcal{A},\mathcal{B}),i\beta  + k}^{} } } } }  \\ 
  + \sum\limits_{k = 1}^\beta  {n_{{\mathcal{R}},k} x_{(\mathcal{A},\mathcal{B}),i\beta  + k}^{} }\,.
 \end{array}
 \end{array}
\end{equation}

The statistical properties of the decision variable $D_{s_{{\mathcal{A}},i} }$ are derived for a fixed bit $i$. The variance of the decision variable is:
\begin{equation}\label{v1}
 \sigma_{D_{s_{{\mathcal{A}},i} } } = {\rm E}\left[ {\left( D_{s_{{\mathcal{A}},i} } \right)^2 } \right] - {\rm E}\left[ D_{s_{{\mathcal{A}},i} } \right]^2  \\
\end{equation}
 Developping \eqref{v1} gives:
\begin{equation}\label{v1_2}
\small
\begin{array}{l}
 \sigma_{D_{s_{{\mathcal{A}},i} } } = {\rm E}\left[ {\left(  \sum\limits_{k = 1}^\beta  {s_{{\mathcal{A}},i} } x_{(\mathcal{A},\mathcal{B}),i\beta  + k}^2  \right)^2 } \right] - \left[   E_b^{(i)}s_{{\mathcal{A}},i}    \right]^2 \\ 
  + {\rm E}\left[  \left( \sum\limits_{k = 1}^\beta  {n_{N,k} x_{(\mathcal{A},\mathcal{B}),i\beta  + k}^{} } \right)^2  \right] \\ 
  + {\rm E}\left[  \left( \sum\limits_{\substack{
        \scriptstyle l \in \mathcal{E} \\ 
        \scriptstyle l \ne m}}\sum\limits_{\scriptstyle m \in \mathcal{E}} {\sum\limits_{k = 1}^\beta  {\left( {s_{l,i}  + s_{m,i} } \right)x_{(l,m),i\beta  + k} x_{(\mathcal{A},\mathcal{B}),i\beta  + k}^{} } }  \right)^2  \right] \\
  
   + {\rm E}\left[ \left( \sum\limits_{k = 1}^\beta  {n_{\mathcal{R},k} x_{(\mathcal{A},\mathcal{B}),i\beta  + k}^{} } \right)^2  \right].
  \end{array}
\end{equation}

For a given fixed bit $i$, the first and the second terms in \eqref{v1_2} are equal since $D_{s_{\mathcal{A},i}}$ is constant. In addition, the spreading sequences and the Gaussian noises are independent and zero mean signals. It follows that the variance of the decision variable is reduced to the variance of the noise terms and the multi-user interference. Therefore, the variance expression becomes:
\begin{equation}\label{v2}
\small
\begin{array}{l}
 \sigma _{D_{s_{{\mathcal{A}},i} } }  =  {\rm E}\left[  \left( \sum\limits_{k = 1}^\beta  {n_{N,k} x_{(\mathcal{A},\mathcal{B}),i\beta  + k}^{} } \right)^2  \right] \\ 
  + {\rm E}\left[  \left(    \sum\limits_{\substack{
        \scriptstyle l \in \mathcal{E} \\
        \scriptstyle l \ne m}}\sum\limits_{\scriptstyle m \in \mathcal{E}} {\sum\limits_{k = 1}^\beta  {\left( {s_{l,i}  + s_{m,i} } \right)x_{(l,m),i\beta  + k} x_{(\mathcal{A},\mathcal{B}),i\beta  + k}^{} } }  \right)^2  \right] \\
  
   + {\rm E}\left[ \left( \sum\limits_{k = 1}^\beta  {n_{\mathcal{R},k} x_{(\mathcal{A},\mathcal{B}),i\beta  + k}^{} } \right)^2  \right].
\end{array}
\end{equation}

The variance of each component in \eqref{v2} is computed as follows.

Since the noise samples are uncorrelated and independent of the chaotic sequence, and the chaotic samples are themselves independent from each other, the conditional variances of the first and third components are:  
\begin{equation} \label{noise1}
\begin{array}{l}
{\rm E}\left[ \left( \sum\limits_{k = 1}^\beta  {n_{N,k} x_{(\mathcal{A},\mathcal{B}),i\beta  + k}^{} } \right)^2  \right] = \beta {\rm E}[ x_{(\mathcal{A},\mathcal{B}),i\beta  + k}^{2}]  N_0/2 \\
 = E_b^{(i)} N_0/2
 \end{array}
\end{equation}
where $E_b^{(i)}=\beta E[x_{(\mathcal{A},\mathcal{B}),i\beta  + k}^{2}]$.

Similarly: 
\begin{equation} \label{noise2}
{\rm E}\left[  \left(    \sum\limits_{k = 1}^\beta  {n_{\mathcal{R},k} x_{(\mathcal{A},\mathcal{B}),i\beta  + k}^{} } \right)^2  \right] = E_b^{(i)} N_0/2.
\end{equation}

Each pair of bits exchanged between $\mathcal{A}$ and $\mathcal{B}$ is spread with the same normalized chaotic sequence. Chaotic sequences are uncorrelated and chaotic samples among a given sequence are independent and normalized with ${\rm E}[x_{k}^2]=1$. It follows that the second term of \eqref{v2} is:
\begin{equation}
\begin{array}{l}
{\rm E}\left[ \left( \left( {s_{l,i}  + s_{m,i} } \right) x_{(l,m),i\beta  + k} x_{(\mathcal{A},\mathcal{B}),i\beta  + k} \right)^2  \right] =\\
2 {\rm E}[ ( x_{(l,m),i\beta  + k} x_{(\mathcal{A},\mathcal{B}),i\beta  + k})^2].
\end{array}
\end{equation}

Then, the  variance of the multi-user interference signal is: 
\begin{equation} \label{mul}
\begin{array}{l}
{\rm E}\left[  \left( \sum\limits_{\substack{
      \scriptstyle l \in \mathcal{E} \\
      \scriptstyle l \ne m}}\sum\limits_{\scriptstyle m \in \mathcal{E}} {\sum\limits_{k = 1}^\beta  {\left( {s_{l,i}  + s_{m,i} } \right)x_{(l,m),i\beta  + k} x_{(\mathcal{A},\mathcal{B}),i\beta  + k}^{} } }  \right)^2  \right]  \\
  =  E_b^{(i)}(L-2).
\end{array}
\end{equation}

Finally, the variance of the decision variable is:
\begin{equation} \label{vr_f}
\begin{array}{l}
 \sigma _{D_{s_{{\mathcal{A}},i} } }   = E_b^{(i)} N_0/2 +E_b^{(i)} N_0/2 + E_b^{(i)} (L-2)\\
=E_b^{(i)} (N_0 + L-2).
\end{array}
\end{equation}

Many papers in the literature apply the Gaussian approach to compute the performance of chaos-based SS communication systems~\cite{Tam04,Fan13}. That approach considers the transmitted bit energy  $E_{b}^{(i)}$ as constant. Given the non-periodic nature of chaotic sequences, the Gaussian approach leads to inaccurate results for small spreading factors\cite{Kad09ieee}. However, for large spreading factors, this approach can be considered as a good approximation that closely matches the exact performance of chaos-based SS communication systems~\cite{Tam04}. In this paper, the Gaussian approach is considered since we are interested in studying our proposed ANC-SS scheme with large spreading factors in order to increase its resistance to multi-user interference. In this case, $E_b^{(i)}$ can be assumed as constant. Thus our BER expression is given by \eqref{ber_e} where the computed variance is from \eqref{vr_f}.

\section{Simulation and Discussion}
\label{sect:sim}

\begin{figure}[htb!]
\centering
% defining custom colors
\definecolor{mycolor1}{rgb}{1,0,1}%
\definecolor{mycolor2}{rgb}{0.168627455830574,0.505882382392883,0.337254911661148}%
\definecolor{mycolor3}{rgb}{0.0431372560560703,0.517647087574005,0.780392169952393}%
\definecolor{mycolor4}{RGB}{154,0,152}% Mauve
\definecolor{mycolor5}{rgb}{0.313725501298904,0.313725501298904,0.313725501298904}%

\begin{tikzpicture}

  \pgfplotsset{
    grid style = {
      dash pattern = on 0.05mm off 1mm,
      line cap = round,
      black,
      line width = 0.5pt
    },
    legend style = {font=\footnotesize},
    tick label style = {font=\footnotesize}
  }

  \begin{semilogyaxis}[
      xmin=0, xmax=15,
      ymin=1e-8, ymax=1,
      xlabel=$E_b/N_0$ (dB),ylabel=BER, ylabel style={yshift=-0.4em}, xlabel style={yshift=0.3em},
      width=\textwidth/2.0, height=8.0cm, grid=major,
      legend style={/tikz/every even column/.append,at={(0.0175,0.0175)},anchor=south west},
      legend cell align=left,
      mark size=2.0pt]

    \addplot [
      color=mycolor1,
      dashed,
      line width=1.0pt,
      mark=x,
      mark options={solid}
    ]
    table[row sep=crcr]{
      0 0.252681666666667\\
      1 0.214674666666667\\
      2 0.177788333333333\\
      3 0.1436455\\
      4 0.112645\\
      5 0.0856976666666667\\
      6 0.0637195\\
      7 0.045968\\
      8 0.0328841666666667\\
      9 0.0233151666666667\\
      10 0.016655\\
      11 0.0120785\\
      12 0.0090015\\
      13 0.0069045\\
      14 0.00551966666666667\\
      15 0.00453283333333333\\
    };
    \addlegendentry{$L=10$, $\beta=100$};

    \addplot [
      color=black,
      solid,
      line width=1.0pt,
      mark=triangle,
      mark options={solid,,rotate=180}
    ]
    table[row sep=crcr]{
      0 0.279501409030535\\
      1 0.244290563312838\\
      2 0.207868692046612\\
      3 0.171637089909259\\
      4 0.137144011481609\\
      5 0.105852305793943\\
      6 0.0788881361140515\\
      7 0.0568527741870733\\
      8 0.039766007597606\\
      9 0.0271566216372379\\
      10 0.0182524381009757\\
      11 0.0121892323545486\\
      12 0.00817063832405572\\
      13 0.00555177894056726\\
      14 0.00385684621952981\\
      15 0.00275763235573212\\
    };
    \addlegendentry{$L=10$, $\beta=100$};

    \addplot [
      color=mycolor4,
      dashed,
      line width=1.0pt,
      mark=asterisk,
      mark options={solid}
    ]
    table[row sep=crcr]{
      0 0.241792222222222\\
      1 0.201355555555556\\
      2 0.161987777777778\\
      3 0.125677777777778\\
      4 0.09192\\
      5 0.0649077777777778\\
      6 0.0420844444444444\\
      7 0.0256055555555556\\
      8 0.0144722222222222\\
      9 0.00725333333333333\\
      10 0.00341555555555556\\
      11 0.00148777777777778\\
      12 0.000595555555555556\\
      13 0.000241111111111111\\
      14 6.88888888888889e-05\\
      15 2.44444444444444e-05\\
    };
    \addlegendentry{$L=10$, $\beta=350$};

    \addplot [
      color=blue,
      solid,
      line width=1.0pt,
      mark=square,
      mark options={solid}
    ]
    table[row sep=crcr]{
      0 0.270685705615462\\
      1 0.232551369037451\\
      2 0.192820649960241\\
      3 0.153188277097065\\
      4 0.115656705537148\\
      5 0.0822331377448839\\
      6 0.0545283163371255\\
      7 0.0333829301927302\\
      8 0.0186836693790108\\
      9 0.00947401412270231\\
      10 0.00432166023216374\\
      11 0.00176635764078082\\
      12 0.000647263747085914\\
      13 0.000214070984481466\\
      14 6.47907249852818e-05\\
      15 1.83298310700156e-05\\
    };
    \addlegendentry{$L=10$, $\beta=350$};

    \addplot [
      color=mycolor2,
      dashed,
      line width=1.0pt,
      mark=star,
      mark options={solid}
    ]
    table[row sep=crcr]{
      0 0.240957714285714\\
      1 0.199752571428571\\
      2 0.159856\\
      3 0.122574857142857\\
      4 0.0893994285714286\\
      5 0.060776\\
      6 0.0383948571428571\\
      7 0.0221634285714286\\
      8 0.0116908571428571\\
      9 0.00541828571428571\\
      10 0.00216914285714286\\
      11 0.000757714285714286\\
      12 0.000238857142857143\\
      13 5.37142857142857e-05\\
      14 1.48571428571429e-05\\
      15 3.42857142857143e-06\\
    };
    \addlegendentry{$L=6$, $\beta=350$};

    \addplot [
      color=mycolor3,
      solid,
      line width=1.0pt,
      mark=triangle,
      mark options={solid}
    ]
    table[row sep=crcr]{
      0 0.268840917373593\\
      1 0.230082299825815\\
      2 0.189645835157144\\
      3 0.149299355495689\\
      4 0.111162700188257\\
      5 0.0773875536182896\\
      6 0.0497125000364973\\
      7 0.0290290086364349\\
      8 0.0151521359758587\\
      9 0.00693986447630931\\
      10 0.0027341980401361\\
      11 0.00090787234856887\\
      12 0.00024913164935159\\
      13 5.55764214458571e-05\\
      14 9.97457647300769e-06\\
      15 1.43881227163041e-06\\
    };
    \addlegendentry{$L= 6$, $\beta=350$};

    \addplot [
      color=red,
      dashed,
      line width=1.0pt,
      mark=x,
      mark options={solid}
    ]
    table[row sep=crcr]{
      0 0.237438285714286\\
      1 0.196086857142857\\
      2 0.155718285714286\\
      3 0.118400571428571\\
      4 0.0847617142857143\\
      5 0.0565914285714286\\
      6 0.0347828571428571\\
      7 0.0191497142857143\\
      8 0.00922342857142857\\
      9 0.00383542857142857\\
      10 0.00124628571428571\\
      11 0.000325714285714286\\
      12 6.51428571428571e-05\\
      13 1.2e-05\\
      14 0\\
      15 0\\
    };
    \addlegendentry{$L=2$, $\beta=350$};

    \addplot [
      color=mycolor5,
      solid,
      line width=1.0pt,
      mark=triangle,
      mark options={solid,,rotate=180}
    ]
    table[row sep=crcr]{
      0 0.266967528662804\\
      1 0.227570679439631\\
      2 0.18641335950309\\
      3 0.145342626731936\\
      4 0.10660716525922\\
      5 0.0725185639992934\\
      6 0.0449556208770412\\
      7 0.0248571994386533\\
      8 0.0119366674882422\\
      9 0.0048149727049047\\
      10 0.00156417701588787\\
      11 0.000387895683388982\\
      12 6.86028943928225e-05\\
      13 7.93846528282e-06\\
      14 5.39029478196876e-07\\
      15 1.87220798061121e-08\\
    };
    \addlegendentry{$L= 2$, $\beta=350$};

    \addplot [
      color=black,
      solid,
      line width=1.0pt
    ]
    table[row sep=crcr]{
      0 0.0786496035251426\\
      1 0.0562819519765415\\
      2 0.037506128358926\\
      3 0.0228784075610853\\
      4 0.0125008180407376\\
      5 0.00595386714777866\\
      6 0.00238829078093281\\
      7 0.000772674815378444\\
      8 0.000190907774075993\\
      9 3.36272284196175e-05\\
      10 3.87210821552204e-06\\
      11 2.6130679535752e-07\\
      12 9.00601035062875e-09\\
      13 1.33293101753005e-10\\
      14 6.81018912878077e-13\\
      15 9.12395736262809e-16\\
    };
    \addlegendentry{BPSK lower bound};

  \end{semilogyaxis}
\end{tikzpicture}%
\caption {Performance of the multi-user ANC-SS scheme. Plain and dashed lines represent analytical and simulation results respectively. $L$ is the number of users and $\beta$ is the spreading factor. The BPSK curve is for a single user.}
\label{anc}
\end{figure}
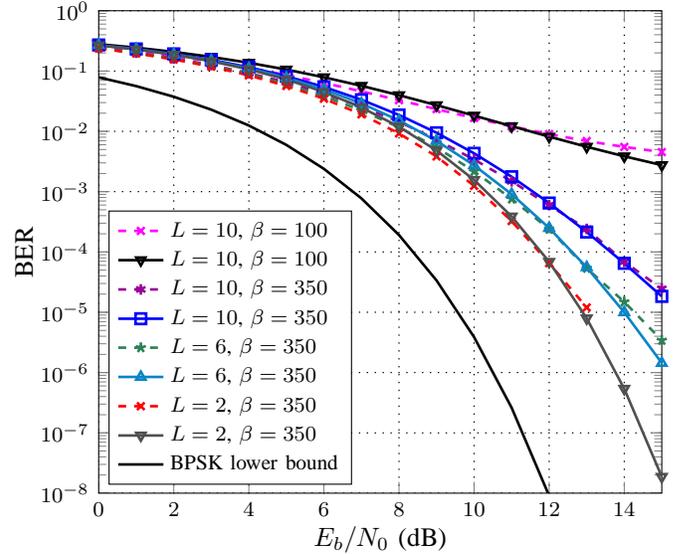

To evaluate the performance of a multi-user ANC-SS system, we plot the computed BER expressions obtained in \eqref{ber_e} against simulation results over an AWGN channel with the presence of multi-user interference. The results shown in Fig.~\ref{anc} are obtained for different numbers of users $L$ and spreading factors $\beta$.

From the figure, it is clear that the computed and simulated BER match closely in all studied cases.

The same figure shows the effect of the spreading factor in the presence of multi-user interference. In such systems, it is suitable to operate with a large spreading factor to reduce the multi-user interference. Indeed, a large spreading factor spreads the interference power over a wide bandwidth, reducing its effect on the demodulated useful signal. 

Fig.~\ref{anc} also plots the BER performance for $L=2$ users. In that case, the network does not have any multi-user interference and the performance can be compared to that of a conventional BPSK modulation.  It can be seen that the performance of a mono-user BPSK point-to-point system is better than the ANC-SS one. The degradation in performance comes from the noise signal added in the transmission phase. This is a disadvantage of the ANC scheme in which the relay does not remove the noise from the received signal; the noise is forwarded to users along with the signal. On the other hand, as explained and shown in this paper, a multi-user ANC-SS system is easier to implement than its physical layer network coding scheme counterpart.

\section{Conclusion}
In this paper, we presented a design for a multi-user ANC-SS system. The goal of the proposed ANC scheme is to reduce the number of time slots required to exchange data, while exploiting interference to increase the throughput. A new coding-spreading method is proposed which consists of having each pair of users that want to communicate between each other use the same spreading code. This leads to a reduction in multi-user interference and also improves the performance of the system. The role of the relay is to synchronize transmission during the first phase and to then broadcast the sum of the signals to all users of the network. The coding and decoding operations are entirely done at the user level without any decoding or mapping in the relay. In our paper, chaotic signals were used as spreading sequences for their good correlation properties. The ANC-SS system was first proposed, then we performed a performance analysis under an AWGN channel and in presence of multi-user interference in order to validate the concept. An analytical BER expression of the ANC-SS was computed and its accuracy confirmed with simulation results.  

\bibliographystyle{IEEEtran}
\bibliography{bibliography}
\end{document}